               \def\pd{\partial}
\def\cs{{\scriptstyle\rm CS}}    
\def\dis{\displaystyle}          
\def\grl{{GR$_\L$}}

       \def\tI{{\tilde I}}
\def\tR{{\tilde R}}   \def\bq{{\bar q}}
\def\bA{{\bar A}}     \def\bu{{\bar u}}
\def\bF{{\bar F}}     \def\bnabla{{\bar\nabla}}

\def\cL{{\cal L}}     \def\cM{{\cal M }}
\def\cO{{\cal O}}     

\def\m{\mu}           \def\n{\nu}           \def\k{\kappa}
                \def\d{\delta}
                
\def\a{\alpha}                 \def\th{\theta}
\def\vphi{\varphi}    \def\ve{\varepsilon}  
\def\r{\rho}                  \def\L{\Lambda}
       \def\om{\omega}

\def\Leff{\hbox{$\L_{\rm eff}$}}
\def\Lra{{\Leftrightarrow}}
\def\bull{\raise.15ex\hbox{\vrule height.8ex width.8ex}}
\def\nn{\nonumber}
\def\be{\begin{equation}}             \def\ee{\end{equation}}
\def\ba#1{\begin{array}{#1}}          \def\ea{\end{array}}
\def\bea{\begin{eqnarray} }           \def\eea{\end{eqnarray} }
\def\beann{\begin{eqnarray*} }        \def\eeann{\end{eqnarray*} }
\def\beal{\begin{eqalign}}            \def\eeal{\end{eqalign}}
\def\lab#1{\label{eq:#1}}             \def\eq#1{(\ref{eq:#1})}
\def\bsubeq{\begin{mathletters}}      \def\esubeq{\end{mathletters}}
\def\bitem{\begin{itemize}}           \def\eitem{\end{itemize}}

\documentstyle[aps,preprint,eqsecnum]{revtex}      

\begin{document}
\tighten


\title{3D gravity with torsion as a Chern-Simons gauge theory}

\author{M.\ Blagojevi\'c and M. Vasili\'c\thanks{Email
        addresses:  mb@phy.bg.ac.yu, mvasilic@phy.bg.ac.yu}}
\address{Institute of Physics, P.O.Box 57, 11001 Belgrade, Serbia}

\maketitle

\begin{abstract}
We show that topological 3D gravity with torsion can be formulated
as a Chern-Simons gauge theory, provided a specific parameter, known
as the effective cosmological constant, is negative. In that case,
the boundary dynamics of the theory corresponding to anti-de Sitter
boundary conditions is described by a conformal field theory with two
different central charges.
\end{abstract}


\section{Introduction}

Investigations of the classical structure of three-dimensional (3D)
Einstein gravity have had an important influence on our understanding
of the related quantum dynamics \cite{1}. In this regard, one should
pay a specific attention to the asymptotic structure of 3D gravity
\cite{2}, as well as to its formulation as a Chern-Simons gauge theory
\cite{3,4}. Following a widely spread belief that the dynamics of
geometry is to be described by general relativity, investigations of 3D
gravity have been carried out mostly in the realm of {\it Riemannian\/}
geometry \cite{5,6,7,8}. Here, we focus our attention on {\it
Riemann-Cartan\/} geometry, in which both the {\it curvature\/} and the
{\it torsion\/} are present as independent geometric characteristics of
spacetime \cite{9,10}.

The asymptotic structure is most clearly seen in topological theories,
in which the non-trivial dynamics can be present only at the boundary
of spacetime. A general action for topological Riemann-Cartan gravity
in 3D has been proposed by Mielke and Baekler (MB)\cite{11,12}. The
model generalizes general relativity with a cosmological constant
(\grl) to a topological gravity in Riemann-Cartan spacetime \cite{13}.
A specific version of the MB model, characterized by the teleparallel
geometry of spacetime without matter, has been recently used to study
the influence of torsion on the asymptotic structure of gravity
\cite{14,15}. The results show that both \grl\ and the teleparallel 3D
gravity have identical asymptotic structures (Chern-Simons formulation,
conformal symmetry and Liouville dynamics at the boundary), at least at
the classical level. Thus, it seems that the asymptotic structure does
not depend on the geometric environment, but only on the boundary
conditions. In the present paper we extend these investigations to the
general MB model in Riemann-Cartan spacetime.

The layout of the paper is as follows. In Sect. II we review the basic
features of the MB model for topological 3D gravity in Riemann-Cartan
spacetime, and analyze the properties that may be of particular
importance for the expected Chern-Simons structure. In Sect. 2 we show
that 3D Riemann-Cartan gravity, in the sector where the so-called
effective cosmological constant is negative, can be formulated as a
Chern-Simons gauge theory. Section IV concludes our exposition.

Our conventions are the same as in Refs. \cite{14,15}: the Latin
indices $(i,j,k,...)$ refer to the local Lorentz frame, the Greek
indices $(\m,\n,\r,...)$ refer to the coordinate frame, and both run
over $0,1,2$; $\eta_{ij}=(+,-,-)$ and
$g_{\m\n}=b^i{_\m}b^j{_\n}\eta_{ij}$ are metric components in the local
Lorentz and coordinate frame, respectively; totally antisymmetric
tensor $\ve^{ijk}$ and the related tensor density $\ve^{\m\n\r}$ are
both normalized by $\ve^{012}=+1$.

\section{Topological 3D gravity with torsion}

In this section we review basic features of the model for topological
Riemann-Cartan gravity in 3D proposed by Mielke and Baekler
\cite{11,12}, and discuss certain dynamical issues which turn out to be
essential for identifying its Chern-Simons structure.

\subsection{Topological gravity in Riemann-Cartan spacetime}

Riemann-Cartan geometry of spacetime can be formulated as Poincar\'e
gauge theory \cite{9,10}: basic gravitational variables are the triad
field $b^i=b^i{_\m}dx^\m$ and the Lorentz connection
$\om^{ij}=\om^{ij}{}_\m dx^\m$ (1-forms), and the related field strengths
are the torsion $T^i$ and the curvature $R^{ij}$ (2-forms). In 3D we
can simplify the notation by introducing the duals of $\om^{ij}$ and
$R^{ij}$:
$$
\om_i=-\frac{1}{2}\,\ve_{ijk}\om^{jk}\, ,\qquad
R_i=-\frac{1}{2}\,\ve_{ijk}R^{jk}\, .
$$
After that, the gauge transformations of the fields take the form
\bsubeq\lab{2.1}
\bea
\d_0 b^i{_\m}&=& -\ve^i{}_{jk}b^j{}_{\m}\th^k
  -\xi^{\r}{}_{,\m}b^i{_\r}-\xi^{\r} b^i{}_{\m,\r}      \nn\\
\d_0\om^i{_\m}&=& -\th^i{}_{,\m}-\ve^i{}_{jk}\om^j{_\m}\th^k
 -\xi^{\r}{}_{,\m}\om^i{_\r}-\xi^{\r}\om^i{}_{\m,\r}\, ,\lab{2.1a}
\eea
and the field strengths are given as (wedge product signs are omitted
for simplicity)
\bea
&&R^i=d\om^i+\frac{1}{2}\,\ve^i{}_{jk}\om^j\om^k
     \equiv\frac{1}{2}R^i{}_{\m\n}dx^\m dx^\n\, ,        \nn\\
&&T^i=db^i+\ve^i{}_{jk}\om^j b^k
     \equiv\frac{1}{2}\,T^i{}_{\m\n}dx^\m dx^\n\, .      \lab{2.1b}
\eea
\esubeq

General gravitational dynamics is defined by demanding the Lagrangian
to be at most quadratic in field strengths \cite{13}. Mielke and
Baekler proposed a {\it topological\/} 3D model, with an action which
is at most {\it linear\/} in field strengths \cite{11,12}:
\bsubeq\lab{2.2}
\be
I=aI_1+\L I_2+\a_3I_3+\a_4I_4+I_M\, ,                   \lab{2.2a}
\ee
where $I_M$ is a matter action, and
\bea
&&I_1=2\int b^iR_i=-\int d^3x\,bR\, ,                         \nn\\
&&I_2=-\frac{1}{3}\,\int\ve_{ijk}b^ib^jb^k=-2\int d^3x\,b\, , \nn\\
&&I_3=\int\left(\om^id\om_i
          +\frac{1}{3}\ve_{ijk}\om^i\om^j\om^k \right)\, ,    \nn\\
&&I_4=\int b^iT_i\, .                                    \lab{2.2b}
\eea
\esubeq
The first two terms are inspired by \grl\ (where $a=1/16\pi G$), $I_3$
is a Chern-Simons action for the Lorentz connection, and $I_4$ is an
action of the translational Chern-Simons type. The MB model can be
thought of as a generalization of \grl\ ($\a_3=\a_4=0$) to a topological
gravity theory in Riemann-Cartan spacetime.

The field equations are obtained by variation with respect to the triad
and connection. In the absence of matter, they take the form
\bea
&&2aR_i+2\a_4T_i-\L\ve_{ijk}b^jb^k=0\,,\nn\\
&&2\a_3R_i+ 2aT_i+\a_4\ve_{ijk}b^jb^k=0\,.\nn
\eea
Assuming $\a_3\a_4-a^2\ne 0$, these equations can be written in the
simple form
\be
2T_i=A\ve_{ijk}b^jb^k\, ,\qquad 2R_i=B\ve_{ijk}b^jb^k\, ,  \lab{2.3}
\ee
where
$$
A=\frac{\a_3\L+\a_4 a}{\a_3\a_4-a^2}\, ,\qquad
B=-\frac{(\a_4)^2+a\L}{\a_3\a_4-a^2}\, .
$$
Thus, the vacuum configuration of fields is characterized by constant
torsion and constant curvature.

In Riemann-Cartan geometry one can use a well known identity to express
the curvature 2-form $R^{ij}$ in terms of its Riemannian piece
$\tR^{ij}$ and the contortion \cite{14}, whereupon the second field
equation in \eq{2.3} can be transformed into an equivalent form,
\be
\tR^{ij}=-\Leff\,b^ib^j\, ,
\qquad \Leff\equiv B-\frac{1}{4}A^2\, ,                    \lab{2.4}
\ee
where $\Leff$ is the effective cosmological constant. Regarded as an
equation for metric, it implies that our spacetime is maximally
symmetric \cite{16}: for $\Leff<0$ ($\Leff>0$) it has the anti-de
Sitter (de Sitter) form. This equation can be considered as an
equivalent of the second equation in \eq{2.3}.

There are two interesting special cases of the general MB model: a) for
$\a_3=\a_4=0$, the Riemann-Cartan theory leads to Riemannian geometry
($A=0$), in the context of which the Chern-Simons structure of gravity
has been first discovered \cite{3}; b) for $\a_3=(\a_4)^2+a\L=0$, the
vacuum geometry becomes teleparallel ($B=0$), but the related
Chern-Simons structure remains the same as in Riemannian case
\cite{14,15}.

\subsection{Riemann-Cartan black hole}

For a given $\Leff$, there is a simple method for finding classical
solutions of the MB theory, defined by the following three-step
procedure: (a) use equation \eq{2.4} to find the metric of our
maximally symmetric space, (b) choose the triad field so as to satisfy
the condition $ds^2=b^ib^j\eta_{ij}$, and (c) use the first equation in
\eq{2.3} to define the related connection $\om^i$.

When the effective cosmological constant is negative,
\be
\Leff\equiv-\frac{1}{\ell^2}<0 \, ,                        \lab{2.5}
\ee
the condition for maximal symmetry \eq{2.4} has a well known solution
for the metric --- the BTZ black hole \cite{17}. Using the static
coordinates $x^\m=(t,r,\vphi)$, and units $4G=1$, it is given as
\bea
&&ds^2=N^2dt^2-N^{-2}dr^2-r^2(d\vphi+N_\vphi dt)^2\, ,\nn\\
&&N^2=\left(-2m+\frac{r^2}{\ell^2}+\frac{J^2}{r^2}\right)\,,
  \qquad N_\vphi=\frac{J}{r^2}\, .                    \nn
\eea
The related triad field can be chosen in the simple form:
\bsubeq\lab{2.6}
\bea
&&b^0=Ndt\, ,\qquad b^1=N^{-1}dr\, ,\nn\\
&&b^2=r\left(d\vphi+N_\vphi dt\right)\, .                \lab{2.6a}
\eea
Then, the connection is obtained by solving the first equation in
\eq{2.3}:
\bea
&&\om^0=N\left(\frac{A}{2}dt-d\vphi\right)\, ,\qquad
  \om^1=N^{-1}\left(\frac{A}{2}+\frac{J}{r^2}\right)dr\,,\nn\\
&&\om^2= -\left(\frac{r}{\ell}
  -\frac{A\ell}{2}\frac{J}{r}\right)\frac{dt}{\ell}
  +\left( \frac{A}{2}r-\frac{J}{r}\right)d\vphi \, .     \lab{2.6b}
\eea
\esubeq
Equations \eq{2.6} define the {\it Riemann-Cartan\/} black hole
\cite{13}. For $A=2/\ell$, and consequently $B=0$, this solution
reduces to the {\it teleparalell\/} black hole \cite{14}.

General solution of the MB model possessing maximal number of {\it
global\/} symmetries defines the Riemann-Cartan AdS solution, AdS$_3$.
It can be obtained from the black hole \eq{2.6} by imposing
$J=0$, $2m=-1$.

In Riemannian geometry, AdS$_3$ is maximally symmetric space, hence it
is locally isometric to any other solution having the same curvature
and signature \cite{16}. Since the same property remains valid in the
MB model, our theory carries no local degrees of freedom.

\subsection{Asymptotic AdS configuration}

Dynamical properties of a theory are very sensitive to the choice of
boundary conditions. They can be thought of as a mechanism for selecting
a class of field configurations which have a special dynamical
importance. Inspired by the results obtained in \grl\ and the
teleparallel 3D gravity, we focus our attention on the {\it asymptotic
AdS configurations\/} of fields, which are determined by the following
requirements \cite{18,2}: (a) they are invariant under the action of
the AdS group, (b) include the black hole configuration \eq{2.6}, and
(c) the related asymptotic symmetries have well defined canonical
generators.

At the moment, we focus our attention on the requirements (a) and (b).
The asymptotics of the triad field $b^i{_\m}$ that satisfies (a) and
(b) is given by the relation
\bsubeq\lab{2.7}
\be
b^i{_\m}= \left( \ba{ccc}
   \dis\frac{r}{\ell}+\cO_1   & O_4  & O_1   \\
   \cO_2 & \dis\frac{\ell}{r}+\cO_3  & O_2   \\
   \cO_1 & \cO_4                     & r+\cO_1
                  \ea
          \right)   \, .                                 \lab{2.7a}
\ee
The arguments leading to this result are the same as those in Ref.
\cite{14}. The asymptotic form of the connection $\om^i{_\m}$ is
defined  with the help of the first equation in \eq{2.3}:
\be
\om^i{_\m}=\left( \ba{ccc}
   \dis\frac{Ar}{2\ell}+\cO_1 & \cO_4 &\dis-\frac{r}{\ell}+\cO_1\\
   \cO_2 & \dis\frac{A\ell}{2r}+\cO_3 & \cO_2  \\
   \dis-\frac{r}{\ell^2}+\cO_1 & \cO_4 & \dis \frac{Ar}{2}+\cO_1
                  \ea
                \right)  \, .                            \lab{2.7b}
\ee
\esubeq
The improved conditions $\om^0{_1},\om^2{_1}=\cO_4$ are in agreement
with the constraints of the theory \cite{14,15}.

A direct verification of the third condition (c) would demand a rather
lengthy canonical analysis. Instead of that, we shall first derive the
Chern-Simons formulation of our theory, whereupon the condition (c)
will follow straightforwardly.

\section{Chern-Simons formulation}

Our experience with \grl\ and the teleparallel 3D theory shows that the
dynamical structure of 3D gravity becomes much simpler if it can be
formulated as an ordinary gauge theory of the Chern-Simons type
\cite{3,15}. In this section we show that the general topological
Riemann-Cartan 3D gravity \eq{2.2} can also be represented as a
Chern-Simons theory.

\subsection{Discovering \boldmath{$SL(2,R)\times SL(2,R)$}
            gauge symmetry}

Every Poincar\'e gauge theory is by construction invariant under the
local Poincar\'e transformations \eq{2.1a}. We shall see that, under
certain conditions, this symmetry can be represented as an
$SL(2,R)\times SL(2,R)$ gauge symmetry.

Let us begin by introducing new variables and gauge parameters by
the relations
\bsubeq
\bea
&&A^i{_\m}=\om^i{_\m}+qb^i{_\m}\, ,\qquad
  \bA^i{_\m}=\om^i{_\m}+\bq b^i{_\m}\, ,                \lab{3.1a}\\
&&u^i=-\th^i-\xi^\m A^i{_\m}\, ,\qquad
  \bu^i=-\th^i-\xi^\m\bA^i{_\m}\, ,                      \lab{3.1b}
\eea
\esubeq
with $q\ne\bq$. Expressed in terms of these, the local Poincar\'e
symmetry \eq{2.1a} takes the form:
\be
\d_0 A^i{_\m}=\nabla_\m u^i+\xi^\r F^i{}_{\m\r}\, ,\qquad
\d_0\bA^i{_\m}=\bnabla_\m\bu^i+\xi^\r\bF^i{}_{\m\r}\, .   \lab{3.2}
\ee
where we use the notation
\bea
&&\nabla_\m u^i=\pd_\m u^i+\ve^i{}_{jk}A^j{_\m}u^k\, ,  \nn\\
&&F^i{}_{\m\n}=\pd_\m A^i{_\n}-\pd_\n A^i{_\m}
                       +\ve^i{}_{jk}A^j{_\m}A^k{_\n}\, ,\nn
\eea
and similarly for $\bnabla$ and $\bF$. Thus, local Poincar\'e symmetry
takes the form of an internal gauge symmetry, provided the field
equations of our theory are equivalent to
\be
F^i{}_{\m\n}=0\, ,\qquad \bF^i{}_{\m\n}=0\, .             \lab{3.3}
\ee
The internal symmetry can be identified with $SL(2,R)\times SL(2,R)$,
as follows from the form of the structure constants $\ve^i{}_{jk}$ and
the metric $\eta_{ij}$ (see, e.g. Ref. \cite{15} for details). Can we
really transform equations \eq{2.3} to the form \eq{3.3}? In order to
verify this, we start from the identities
\bea
&&F^i{}_{\m\n}=R^i{}_{\m\n}+qT^i{}_{\m\n}
     +q^2\ve^i{}_{jk}b^j{_\m}b^k{_\n}\, ,           \nn\\
&&\bF^i{}_{\m\n}=R^i{}_{\m\n}+\bq T^i{}_{\m\n}
     +\bq^2\ve^i{}_{jk}b^j{_\m}b^k{_\n}\, ,         \nn
\eea
which imply that equations \eq{3.3} can be rewritten as
\be
T_{ijk}=-(q+\bq)\ve_{ijk}\, ,\qquad
R_{ijk}=q\bq\ve_{ijk}\, .                                  \lab{3.4}
\ee
Consequently, equations \eq{3.3} coincide with \eq{2.3}
if $A=-(q+\bq)$, $B=q\bq$, or equivalently:
\bea
&&2q=-A + \sqrt{A^2-4B}\, ,  \nn\\
&&2\bq=-A - \sqrt{A^2-4B}\, .                              \lab{3.5}
\eea
Demanding that the parameters $q$ and $\bq$ be real and different
from each other, we obtain the following restrictions on $A$ and $B$:
\be
A^2-4B>0 \quad\Lra\quad \Leff\equiv B-\frac{1}{4}A^2<0\, . \lab{3.6}
\ee
\bitem
\item[--] In the AdS sector of the MB theory, i.e. for $\Leff<0$,
the gravitational field equations are equivalent to the Chern-Simons
equations \eq{3.3}, and local Poincar\'e symmetry coincides (on shell)
with the $SL(2,R)\times SL(2,R)$ gauge symmetry.
\eitem

\subsection{Chern-Simons form of the action}

Having found two independent $SL(2,R)$ gauge symmetries at the level of
field equations, we now wish to find out, following Witten \cite{3},
whether the action of 3D gravity can be represented as a combination of
two pieces, each of which depends only on one of two independent gauge
fields, $A$ or $\bA$.

Starting from the Chern-Simons Lagrangian,
\be
\cL_\cs(A)=A^idA_i +\frac{1}{3}\ve_{ijk}A^iA^jA^k \,,      \lab{3.7}
\ee
where $A^i=A^i{_\m}dx^\m$, $A_i=\eta_{ij}A^j$, we can use the
expressions \eq{3.1a} for $A^i$ i $\bA^i$ and obtain the relation
$$
\cL_\cs(A)=\cL_\cs(\om)+2qb^iR_i+q^2b^iT_i
     +\frac{1}{3}q^3\ve_{ijk}b^ib^jb^k+qd(b^i\om_i)\, .
$$
This result leads directly to the important identity
\bsubeq
\bea
\k_1\cL_\cs(A)-\k_2\cL_\cs(\bA)=&&
    2a b^iR_i -\frac{1}{3}\L\ve_{ijk}b^ib^jb^k           \nn\\
  &&+\a_3\cL_\cs(\om)+\a_4 b^iT_i + ad(b^i\om_i)\, ,     \lab{3.8a}
\eea
where
\bea
&&a=\k_1q-\k_2\bq\, ,\quad\,\; \L=-(\k_1q^3-\k_2\bq^3)\, , \nn\\
&&\a_3=\k_1-\k_2\, ,\qquad
  \a_4=\k_1q^2-\k_2\bq^2\, .                             \lab{3.8b}
\eea
\esubeq
Comparing with \eq{2.2} we obtain the final result:
\be
\k_1\cL_\cs(A)-\k_2\cL_\cs(\bA)=\cL_G+ ad(b^i\om_i)\, ,   \lab{3.9}
\ee
where $\cL_G$ denotes the gravitational Lagrangian in \eq{2.2}.

Let us note that under the boundary conditions \eq{2.7} the
Chern-Simons actions $I_\cs[A]$ and $I_\cs[\bA]$ do not have well
defined functional derivatives \cite{19,20}. However, without loss of
generality, these conditions can be refined to ensure the needed
differentiability, as we shall see in the next subsection. After that,
equation \eq{3.9} can be written in the simple form
\bsubeq\lab{3.10}
\be
\k_1I_\cs[A]-\k_2I_\cs[\bA]=\tI_G\, ,                    \lab{3.10a}
\ee
where $\tI_G$ is the improved (differentiable) gravitational MB action:
\be
\tI_G=I_G+ a\int d(b^i\om_i)\, .                         \lab{3.10b}
\ee
\esubeq
\bitem
\item[--] The improved gravitational action of the Riemann-Cartan
3D gravity is equal to a linear combination of two Chern-Simons
actions.
\eitem
The role of coefficients $\k_1$ and $\k_2$ is to define central charges
of the theory \cite{5}.

The parameters $(a,\L,\a_3,\a_4)$ are functionally independent of
$(\k_1,\k_2,q,\bq)$ if the Jacobian of the transformations does not
vanish. From
$$
{\rm det\,}\frac{\pd(a,\L,\a_3,\a_4)}{\pd(\k_1,\k_2,q,\bq)}
      =-\k_1\k_2(q-\bq)^4\, ,
$$
we see that the transformation of parameters is regular for
$q\ne\bq$. Using equations \eq{3.8b}, we can explicitly express
$(\k_1,\k_2,q,\bq)$ in terms of $(a,\L,\a_3,\a_4)$:
\bea
&&q=-\frac{A}{2}+\frac{1}{\ell}\, ,\qquad
    \bq=-\frac{A}{2}-\frac{1}{\ell}\, ,                 \nn\\
&&\k_1-\k_2=\a_3\, ,\qquad
    \k_1+\k_2=\ell\left(a+\frac{A}{2}\,\a_3\right)\, .  \lab{3.11}
\eea

The above relation clarifies the role of four parameters appearing in
the action \eq{2.2}. In particular, our gravitational theory with
$\a_3\ne 0$ (and boundary conditions \eq{2.7}) has conformal symmetry
with {\it two different central charges\/}:
\be
c_i=12\cdot 4\pi\k_i \qquad (i=1,2)\, .                   \lab{3.12}
\ee
 Table 1 illustrates two particular cases belonging to the
complementary sector $\a_3=0$.

\def\rl{\rule[-1.2ex]{0pt}{3.5ex}}
\begin{center}
\doublerulesep 1.5pt
\begin{tabular}{|l|l|c|c|c|c|}
\multicolumn{6}{l}{\rl Table 1. Particular cases of the
                       MB model with $\Leff<0$}  \\   \hline\hline
\rl ~Conditions   &~Geometry
     &~$q$~&~$\bq$~&~$\k_1$~&~$\k_2~$~           \\   \hline\hline
\rule[-1pt]{0pt}{15pt}
\rl  $\a_3=0\,,\,\, A=0~$ &~Riemannian
     &~$1/\ell$~&~$-1/\ell~$ &~$a\ell/2$~&~$a\ell/2$~\\
\rule[-1pt]{0pt}{15pt}
\rl  $\a_3=0\,,\,\, B=0~$ &~Teleparallel
     &~$0$~&~$-2/\ell$~&~$a\ell/2$~&~$a\ell/2$~  \\   \hline\hline
\end{tabular}
\end{center}
\smallskip
The first case corresponds to \grl\ \cite{3}, the second one to the
teleparallel theory \cite{14}. In both cases we have two copies of the
same central charge,
$$
c_1=c_2=48\pi\frac{a\ell}{2}=\frac{3\ell}{2G}\, .
$$

We can now return to the condition (c) formulated in subsection II.C.
Using the relations \eq{3.10} and the known results concerning the
canonical structure of the Chern-Simons theory, see e.g. Refs.
\cite{5,10}, one can conclude that the asymptotic symmetries of the
theory defined by the action $\tI_G$ and the boundary conditions
\eq{2.7} have well defined canonical generators.

\subsection{Asymptotic conditions}

Since the dynamical content of Chern-Simons theory is influenced by the
form of boundary conditions, we shall now investigate the corresponding
behavior of $A$ and $\bA$.

The asymptotic conditions \eq{2.7} for the gravitational variables
$\om^i$ and $b^i$, in conjunction with the values of $q$ and $\bq$ given
in \eq{3.11}, lead to the following conditions on $A^i$ and $\bA^i$:
\bsubeq\lab{3.13}
\be
A^i{_\m}=\left( \ba{ccc}
  \dis\frac{r}{\ell^2}+\cO_1 & \cO_4 &\dis -\frac{r}{\ell}+\cO_1\\
  \cO_2 & \dis\frac{1}{r}+\cO_3 & \cO_2   \\
  \dis-\frac{r}{\ell^2}+\cO_1 & \cO_4 & \dis \frac{r}{\ell}+\cO_1
                 \ea
          \right) \, ,                                  \lab{3.13a}
\ee
\be
\bA^i{_\m}=\left( \ba{ccc}
  \dis-\frac{r}{\ell^2}+\cO_1 & \cO_4 &\dis -\frac{r}{\ell}+\cO_1\\
  \cO_2 & \dis-\frac{1}{r}+\cO_3 & \cO_2   \\
  \dis-\frac{r}{\ell^2}+\cO_1 & \cO_4 & \dis-\frac{r}{\ell}+\cO_1
                   \ea
            \right) \, .                                 \lab{3.13b}
\ee
\esubeq
Using the light-cone notation, $x^\pm=t/\ell\pm\vphi$,
$A^\pm=A^0\pm A^2$, the conditions for $A^i$ can be rewritten in the
form:
\bsubeq\lab{3.14}
\bea
&&A^1=\frac{dr}{r}+\cO_2\, ,\qquad A^+=\cO_1\, ,         \nn\\
&&A^-=\frac{2r}{\ell}dx^- +\cO_1\, ,                     \lab{3.14a}
\eea
with the additional restrictions on $A_1$:
\be
A^\pm_1=\cO_4\, ,\qquad A^1_1=\frac{1}{r}+\cO_3\, .      \lab{3.14b}
\ee
\esubeq
Similarly, we find for $\bA^i$:
\bsubeq\lab{3.15}
\bea
&&\bA^1=-\frac{dr}{r}+\cO_2\, ,\qquad \bA^-=\cO_1\, ,    \nn\\
&&\bA^+=-\frac{2r}{\ell}dx^+ +\cO_1\, ,                  \lab{3.15a}
\eea
together with
\be
\bA^\pm_1=\cO_4\, ,\qquad \bA^1_1=-\frac{1}{r}+\cO_3\, . \lab{3.15b}
\ee
\esubeq

Having found the boundary conditions for $A$ and $\bA$, we now return
to the interpretation of the action $\tI_G$ in \eq{3.10}. The variation
of the Chern-Simons action takes the form
$$
\d I_\cs[A]= \int_{\cM}\d A^iF_i+\int_{\pd\cM}A^i\d A_i\, ,
$$
hence it has well defined functional derivatives if the boundary term
vanishes. Relations \eq{3.14} and \eq{3.15} yield the conditions on
$A_+$ and $\bA_-$,
\bea
&& A^1_+=\cO_2\, ,\quad A^-_+=\cO_1\, ,\quad A^+_+=\cO_1\, ,\nn\\
&&\bA^1_-=\cO_2\, ,\quad \bA^+_-=\cO_1\,,\quad \bA^-_-=\cO_1\,,\nn
\eea
which are seen to be insufficient for the differentiability of
$I_\cs[A]$ and $I_\cs[\bA]$. Fortunately, we can get rid of the problem
in a simple manner --- by adopting the refined conditions
$$
A^+_+=\cO_3\, ,\qquad \bA^-_-=\cO_3\, ,
$$
compatible with the field equations. Equation \eq{3.10a} then implies
that $\tI_G$ is also differentiable.

The boundary conditions \eq{3.14} and \eq{3.15} have the same form as
in the teleparallel case \cite{15}, the only difference being the
different value of the constant $\ell$. Following the same line of
arguments, based on the form of field equations and boundary
conditions, one verifies that the complete dynamics is located at the
boundary, and is described by two chiral fields (functions of only
$x^+$ or $x^-$). According to the analysis of the preceding subsection,
\bitem
\item[--] general dynamical evolution of these boundary fields follows
the rules of a conformal field theory possessing two different
classical central charges.
\eitem

\section{Concluding remarks}

In this paper we show that the general Riemann-Cartan topological 3D
gravity \eq{2.2} can be formulated as an $SL(2,R)\times SL(2,R)$
Chern-Simons gauge theory, provided we stay in its AdS sector, defined
by the condition $\Leff<0$. More precisely, we show that
\bitem
\item[i)] the improved gravitational action \eq{2.2} is represented
as a linear combination of two independent Chern-Simons actions,
equation \eq{3.10},
\eitem
and also that the gravitational AdS boundary conditions \eq{2.7}
transform into the standard Chern-Simons boundary conditions \eq{3.13}.
As a consequence,
\bitem
\item[ii)] boundary dynamics of the general Riemann-Cartan
gravity \eq{2.2} is described by a conformal field theory with two
different classical central charges,
\eitem
in contrast to the cases of \grl\ and the teleparallel theory
\cite{3,14}.

One should note that in \grl\ and the teleparallel gravity, where
$c_1=c_2$, the non-trivial boundary components of $A$ and $\bA$ are
associated with a single Liouville field \cite{4,14}. The
corresponding effective two-dimensional action (the Liouville
action) correctly reproduces both the field equations and the
central charge of the associated gravity theory.
In the general MB theory, the boundary fields can still be
associated with a single Liouville field subject to the Liouville
field equations. The Liouville action, however, does not correctly
reproduce the needed two central charges. It would be interesting to
find the form of an effective 2-dimensional action that would
correctly reproduce both the field equations and the central charges
of the general MB theory.

\acknowledgements
This work was partially supported by the Serbian Science foundation,
Serbia.

\end{document}